\documentstyle[amssymb,pra,aps,epsfig,12pt]{revtex}

\newcommand{\be}{\begin{equation}}
\newcommand{\ee}{\end{equation}}
\newcommand{\br}{\begin{eqnarray}}
\newcommand{\er}{\end{eqnarray}}

\newcommand{\bd}{\begin{displaymath}}
\newcommand{\ed}{\end{displaymath}}

\newcommand{\bfig}{\begin{figure}}
\newcommand{\efig}{\end{figure}}

\def\3cdot{\cdot \cdot \cdot}

\def\om0{\omega _0}
\def\Om0{\Omega _0}

\def\text#1{{\rm{#1}}}

\def\->{\rightarrow}
\def\=>{\Rightarrow}
\def\-->{\longrightarrow}
\def\==>{\Longrightarrow}

\def\pr{^\prime}

\def\pr2{^{\prime\prime}}

\def\bfig{\begin{figure}}
\def\efig{\end{figure}}

\begin{document}
\title{{\Large Single-particle quantum tunneling in ionic traps}}
\author{R. M. Serra\thanks{%
E-mail: serra@df.ufscar.br}, C. J. Villas-B\^{o}as, and M. H. Y. Moussa%
\thanks{%
E-mail: miled@df.ufscar.br}}
\address{Departamento de F\'{\i}sica, Universidade Federal de S\~{a}o\\
Carlos, P.O. Box 676, S\~{a}o Carlos, 13565-905, S\~{a}o Paulo, Brazil. }
\maketitle

\begin{abstract}
We describe a proposal to probe the quantum tunneling mechanism of an
individual ion trapped in a double-well electromagnetic potential. The
time-evolution of the probability of fluorescence measurement of the
electronic ground state is employed to characterize the single-particle
tunneling mechanism. The proposed scheme can be used to implement quantum
information devices.

{\bf Journal Ref.} J. Opt. B: Quantum Semiclass. Opt. {\bf 5}, 237-242 (2003)


\end{abstract}

Together with nonlocality and wave-vector collapse -- the phenomena behind
the recent advances in quantum communication \cite{comm} and computation %
\cite{comp} -- the tunneling mechanism is one of the most intriguing aspects
of the microscopic world of particles and their interactions, setting
quantum reality apart from classical physics. While nonlocality and
wave-vector collapse have been pursued as clues to understanding fundamental
quantum phenomena such as the uncertainty principle and the process of
quantum measurement \cite{EPR}, quantum tunneling is a valuable mechanism
for probing the transition from quantum to classical dynamics \cite{CL}.

Quantum tunneling at the macroscopic level has attracted much attention in
the literature \cite{CL,Leggett,Milburn}, firstly owing to its application
to SQUIDs (superconducting quantum interference devices), permitted by the
advances in cryogenics, and recently because of the realization of
Bose-Einstein condensation in dilute atomic gases. The observation of
matter-wave interference fringes has demonstrated that a Bose condensate
consists of ``laser-like'' atoms which are spatially coherent and show
long-range correlations, opening the field of coherent atomic beams and of
atomic Josephson effect \cite{Ketterle}.

Probed at the single-particle level, quantum tunneling is just as
provocative to our intuition as its manifestation on the macroscopic scale.
The experimental techniques developed over the last decade for manipulating
electronic and vibrational states of trapped ions can be employed to
investigate fundamental quantum effects at a level so far accessible only as
collective processes \cite{Wineland}. We describe in this letter a scheme
for probing the tunneling mechanism of an individual ion trapped in a
double-well electromagnetic potential. The time evolution of the probability
of a fluorescence measurement of the electronic ground state is used to
probe the single-particle tunneling mechanism.

We first assume $i)$ that the ion is trapped in a symmetric double-well
single-particle potential $V(x)=b\left( x^{2}-x_{0}^{2}\right) ^{2}$. (In
ionic traps, the typical oscillation frequencies in the $y$ and $z$
directions are very much larger than that in the $x$ direction, providing a
good approximation to a one-dimensional trap.) The minima are given by $%
x_{0}=\pm \sqrt{d/2b}$, so that the motion can be described as approximately
harmonic, with frequency $\omega _{0}=\sqrt{4d/m}$, $m$ being the ionic
mass. The harmonic approximation can be adjusted by fixing $d$ (and
consequently $\omega _{0}$) and conveniently choosing the parameter $b$,
which is equivalent to varying the height of the barrier separating the two
wells, $h=d^{2}/4b$. We also assume $ii)$ that the parabolic approximation
to the potential around each minimum is designed to contain (at least) the
two lowest states of the harmonic oscillator described by the wave functions 
$\phi _{{\bf i}}^{\left( n\right) }\left[ x-(-1)^{{\bf i}}x_{0}\right] $,
with $n=1,2$ referring to the ground and first-excited states of the
harmonic oscillator and ${\bf i}=1,2$ referring to both harmonic wells
centered at $x=\mp x_{0}$. Finally, we assume $iii)$ that the potential is
such that state $\phi _{1}^{\left( n\right) }$ (of harmonic trap $1$) is
spatially close to\ state $\phi _{2}^{\left( n\right) }$ (of harmonic trap $%
2 $). Thus, these local modes are not orthogonal, due to the overlap $%
\epsilon $ between the corresponding modes of the two wells: 
\begin{equation}
\int dx\left( \phi _{{\bf i}}^{\left( n\right) }\left[ x-(-1)^{{\bf i}}x_{0}%
\right] \right) ^{\ast }\phi _{{\bf j}}^{\left( m\right) }\left[ x-(-1)^{%
{\bf j}}x_{0}\right] =\left[ \delta _{{\bf ij}}+\epsilon \left( 1-\delta _{%
{\bf ij}}\right) \right] \delta _{nm}{.}  \label{Eq1}
\end{equation}
When $\epsilon \ll 1$, given a first order correction , these local modes
are approximately orthogonal and the eigenstates of the global double-well
potential may be approximated by the symmetric and asymmetric superpositions 
$\phi _{\pm }^{\left( n\right) }(x)\approx \left[ \phi _{1}^{\left( n\right)
}\left( x+x_{0}\right) \pm \phi _{2}^{\left( n\right) }\left( x-x_{0}\right) %
\right] /\sqrt{2}$, with corresponding eigenvalues $E_{\pm }^{\left(
n\right) }=E_{0}^{\left( n\right) }\pm {\cal R}^{\left( n\right) }$. The
coupling constants between the local states are given by the matrix elements 
\begin{equation}
{\cal R}^{\left( n\right) }=\int dx\left[ \phi _{1}^{\left( n\right) }\left(
x+x_{0}\right) \right] ^{\ast }\left[ V(x)-\widetilde{V}_{1}\left[ x+x_{0}%
\right] \right] \phi _{2}^{\left( n\right) }\left( x-x_{0}\right) ,
\label{Eq2}
\end{equation}
where $\widetilde{V}_{{\bf i}}\left[ x-(-1)^{{\bf i}}x_{0}\right] $ indicate
the parabolic potential approximations around $x=(-1)^{{\bf i}}x_{0}$. The
tunneling frequencies $\Omega ^{\left( n\right) }$ between the corresponding
eigenstates in $\widetilde{V}_{1}$ and $\widetilde{V}_{2}$ are given by $%
\Omega ^{\left( n\right) }=2\left| {\cal R}^{\left( n\right) }\right| /\hbar 
$. In order to make the first order correction ($\epsilon ^{1}$) valid,
ensuring that ${\cal R}^{\left( n\right) }/E_{0}^{\left( n\right) }\ll 1$,
we further impose the condition $x_{0}^{2}/\Delta _{x}^{2}\gg 1$. Scaling
the length in units of the position uncertainty in a harmonic oscillator
ground state $\Delta _{x}^{2}=\hbar /\left( 2m\omega _{0}\right) $, we
obtain from Eq. (\ref{Eq2}) $\Omega ^{\left( 1\right) }=\left( \left.
3\omega _{0}x_{0}^{2}\right/ 8\Delta _{x}^{2}\right) \exp \left(
-x_{0}^{2}/2\Delta _{x}^{2}\right) $ and $\Omega ^{\left( 2\right) }=\left(
x_{0}^{2}/\Delta _{x}^{2}\right) $ $\Omega ^{\left( 1\right) }$.

The Hamiltonian describing the motional degrees of freedom of the ion
trapped in the above-described double-well potential is 
\begin{equation}
\widehat{H}_{motional}=\int dx\widehat{\Psi }^{\dagger }H\widehat{\Psi }{\rm 
{,}\quad }H=-\frac{\hbar ^{2}}{2m}\nabla ^{2}+V\left( x\right) ,  \label{Eq3}
\end{equation}
where $\widehat{\Psi }\left( x\right) =\sum_{n,{\bf i}}\phi _{{\bf i}%
}^{\left( n\right) }\left( x\right) \widehat{c}_{{\bf i}}^{\left( n\right) }$
is the field operator which annihilates the ion in the state $\phi _{{\bf i}%
}^{\left( n\right) }$, i.e., in energy level $E_{0}^{\left( n\right) }$ of
the harmonic well $\widetilde{V}_{{\bf i}}$. From the eigenvalue equation $%
H\phi _{\pm }^{\left( n\right) }=E_{\pm }^{\left( n\right) }\phi _{\pm
}^{\left( n\right) }$ and the assumption that $\epsilon \ll 1$, the
Hamiltonian in Eq. (\ref{Eq3}) becomes

\begin{equation}
\widehat{H}_{motional}=\sum\nolimits_{n}\left[ E_{0}^{\left( n\right)
}\left( \widehat{c}_{1}^{\left( n\right) ^{\dagger }}\widehat{c}_{1}^{\left(
n\right) }+\widehat{c}_{2}^{\left( n\right) ^{\dagger }}\widehat{c}%
_{2}^{\left( n\right) }\right) +{\cal R}^{\left( n\right) }\left( \widehat{c}%
_{1}^{\left( n\right) ^{\dagger }}\widehat{c}_{2}^{\left( n\right) }+%
\widehat{c}_{2}^{\left( n\right) ^{\dagger }}\widehat{c}_{1}^{\left(
n\right) }\right) \right] .  \label{Eq4}
\end{equation}

Next, we assume that the trapped ion has two effective electronic states,
excited $\left| \uparrow \right\rangle $ and ground $\left| \downarrow
\right\rangle $, separated by frequency $\omega $ and coupled by the
interaction with an effective laser plane wave propagating in the $x$
direction, with wave vector $k_{L}=\omega _{L}/c$. In this configuration,
only the ionic motion along the $x$\ axis will be modified.{\large \ }The
effective pumping laser beam, is detuned by $\delta \equiv \omega -\omega
_{L}$ from the $\left| \uparrow \right\rangle $ $\leftrightarrow $ $\left|
\downarrow \right\rangle $ transition. In NIST experiments with $^{9}$Be$%
^{+} $ the laser beams (with different frequencies) composing the effective
pumping laser, are detuned from a third more excited level which, in the
stimulated Raman-type configuration, is adiabatically eliminated \cite%
{Wineland}. In the Innsbruck experiments with $^{40}$Ca$^{+}$, a coherently
driving direct transition between the electronic ground and excited states
is employed \cite{Blatt}. A {\it probe} laser, strongly coupled to the
transition between the electronic ground state and a third more excited
level $|s\rangle $, is also employed in order to measure the ionic
vibrational state by collecting the resonance fluorescence signal, which is
the probability of the ion being found in the internal state{\large \ }$%
\left| \downarrow \right\rangle $.

We can trail the tunneling phenomenon through the time-evolution of the
probability of fluorescence measurement of the electronic ground state
collected in two different situations: $a)$ by individual addressing the ion
in harmonic well $1$ or, what seems to be more attractive experimentally, $%
b) $ considering the laser beams (the pumping and the probe lasers) to reach
both local potential wells simultaneously. The ion-laser interaction
Hamiltonian that describes the effective interaction of the quantized motion
of the ionic center-of-mass coupled to its electronic degrees of freedom is %
\cite{Wineland} 
\begin{equation}
\widehat{H}_{ion-laser}=\hslash g\left( \widehat{\sigma }_{+}%
\mathop{\rm e}%
\nolimits^{ik_{L}\widehat{x}-i\omega _{L}t+i\varphi }+\widehat{\sigma }_{-}%
\mathop{\rm e}%
\nolimits^{-ik_{L}\widehat{x}+i\omega _{L}t-i\varphi }\right) ,  \label{Eq5}
\end{equation}
where $\sigma _{+}=\left| \uparrow \right\rangle \left\langle \downarrow
\right| $ and $\sigma _{-}=\left| \downarrow \right\rangle \left\langle
\uparrow \right| $ are the usual Pauli pseudo-spin operators, $g$ is the
effective Rabi frequency of the transition $\left| \uparrow \right\rangle
\leftrightarrow \left| \downarrow \right\rangle $ and $\phi $ is the phase
difference between the two lasers composing the effective beam. Since,
considering the situation $a)$, the laser beams are disposed in harmonic
well $1$, the position operator $\widehat{x}$ is given by

\begin{eqnarray}
\widehat{x} &=&\sum\nolimits_{n,m}\int dx\phi _{1}^{\left( n\right)
^{\dagger }}\left( x\right) x\phi _{1}^{\left( m\right) }\left( x\right) 
\widehat{c}_{1}^{\left( n\right) ^{\dagger }}\widehat{c}_{1}^{\left(
m\right) }  \nonumber \\
&=&\Delta _{x}\sum\nolimits_{n}\sqrt{n+1}\left( \widehat{c}_{1}^{\left(
n\right) ^{\dagger }}\widehat{c}_{1}^{\left( n+1\right) }+\widehat{c}%
_{1}^{\left( n+1\right) ^{\dagger }}\widehat{c}_{1}^{\left( n\right)
}\right) .  \label{Eq6}
\end{eqnarray}
Evidently, considering the situation $b)$ where the laser beams reach both
local wells, the position operator turn to be $\widehat{x}=\Delta
_{x}\sum\nolimits_{j=1,2}\sum\nolimits_{n}\sqrt{n+1}\left( \widehat{c}%
_{j}^{\left( n\right) ^{\dagger }}\widehat{c}_{j}^{\left( n+1\right) }+%
\widehat{c}_{j}^{\left( n+1\right) ^{\dagger }}\widehat{c}_{j}^{\left(
n\right) }\right) $.

In what follows we consider only the situation $a)$ since the generalization
for the situation $b)$, which will be analyzed further, is straightforward.
In a frame rotating at the effective laser frequency $\omega _{L}$, the
ion-laser Hamiltonian is given by 
\begin{equation}
\widehat{H}_{ion-laser}=\hbar g\left[ \widehat{\sigma }_{+}\exp \left( i\eta
\sum\nolimits_{n}\sqrt{n+1}\left( \widehat{c}_{1}^{\left( n\right) ^{\dagger
}}\widehat{c}_{1}^{\left( n+1\right) }+\widehat{c}_{1}^{\left( n+1\right)
^{\dagger }}\widehat{c}_{1}^{\left( n\right) }\right) +i\varphi _{L}\right) +%
{\rm {H.c.}}\right] ,  \label{Eq7}
\end{equation}
where we have introduced the Lamb-Dicke parameter $\eta \equiv k_{L}\Delta
_{x}$.

In what follows we rewrite the total Hamiltonian in the interaction picture
(script font labels) via the unitary transformation $\widehat{U}(t)=\exp
\left( -i\widehat{H}_{0}t\right) $, where $\widehat{H}_{0}=\sum%
\nolimits_{n}E_{0}^{\left( n\right) }\left( \widehat{c}_{1}^{\left( n\right)
^{\dagger }}\widehat{c}_{1}^{\left( n\right) }+\widehat{c}_{2}^{\left(
n\right) ^{\dagger }}\widehat{c}_{2}^{\left( n\right) }\right) +\hbar \frac{%
\delta }{2}\widehat{\sigma }_{z}$ indicates the free Hamiltonian composed of
the internal and motional degrees of freedom of the trapped ion, $\sigma
_{z} $ being the Pauli operator. The resulting Hamiltonian, including the
motional and electronic degrees of freedom, reads 
\begin{eqnarray}
\widehat{{\cal H}} &=&\sum\nolimits_{n}{\cal R}^{\left( n\right) }\left( 
\widehat{c}_{1}^{\left( n\right) ^{\dagger }}\widehat{c}_{2}^{\left(
n\right) }+\widehat{c}_{2}^{\left( n\right) ^{\dagger }}\widehat{c}%
_{1}^{\left( n\right) }\right) +\hbar g\exp \left( -\eta
^{2}\sum\nolimits_{n}\widehat{c}_{1}^{\left( n\right) ^{\dagger }}\widehat{c}%
_{1}^{\left( n\right) }\right)  \nonumber \\
&&\times \left\{ \sigma _{+}\exp \left[ i\eta \sum\nolimits_{n}\sqrt{n+1}%
\left( \widehat{c}_{1}^{\left( n\right) ^{\dagger }}\widehat{c}_{1}^{\left(
n+1\right) }e^{-i\omega _{0}t}+\widehat{c}_{1}^{\left( n+1\right) ^{\dagger
}}\widehat{c}_{1}^{\left( n\right) }e^{i\omega _{0}t}\right) \right]
e^{i\delta t+i\varphi _{L}}+{\rm {H.c.}}\right\}  \label{Eq8}
\end{eqnarray}
In order to simplify the expression (\ref{Eq8}) we $i)$ adjust the ion-laser
detuning to the first red sideband for the ion-laser interaction ($\delta
=\omega _{0}$). In addition, $ii)$ the assumption of the standard Lamb-Dicke
limit ($\eta \ll 1$), where the ionic center of mass is strongly localized
with respect to the laser wavelength, enables the expansion of the
Hamiltonian (\ref{Eq8}) to first order correction ($\eta ^{1}$). Finally, $%
iii)$ with the optical rotating wave approximation, we obtain the expression 
\begin{eqnarray}
\widehat{{\cal H}} &=&\sum\nolimits_{n}{\cal R}^{\left( n\right) }\left( 
\widehat{c}_{1}^{\left( n\right) ^{\dagger }}\widehat{c}_{2}^{\left(
n\right) }+\widehat{c}_{2}^{\left( n\right) ^{\dagger }}\widehat{c}%
_{1}^{\left( n\right) }\right)  \nonumber \\
&&+i\hbar \eta g\exp \left( -\eta ^{2}\sum\nolimits_{n}\widehat{c}%
_{1}^{\left( n\right) ^{\dagger }}\widehat{c}_{1}^{\left( n\right) }\right)
\left( \sigma _{+}\sum\nolimits_{n}\sqrt{n+1}\widehat{c}_{1}^{\left(
n\right) ^{\dagger }}\widehat{c}_{1}^{\left( n+1\right) }e^{+i\varphi _{L}}-%
{\rm {H.c.}}\right) {\rm {,}}  \label{Eq9}
\end{eqnarray}
which in the absence of the tunneling mechanism leads to the coupling
between the electronic and motional degrees of freedom of the trapped ion,
described by the Jaynes-Cummings Hamiltonian (JCH).

At this point, we assume that the ion is initially cooled to its motional
ground state in harmonic well $1$, and laser-excited to the electronic state 
$\left| \uparrow \right\rangle $. Thus, the JCH induces the transition $%
\left| 1,0\right\rangle _{1}$ $\left| \uparrow \right\rangle \leftrightarrow
\left| 0,1\right\rangle _{1}$ $\left| \downarrow \right\rangle $, where the
ket $\left| 1,0\right\rangle _{{\bf l}}$ ($\left| 0,1\right\rangle _{{\bf l}%
} $) indicates the ion in the motional ground (excited) state of $\widetilde{%
V}_{{\bf l}}$. Evidently, being in the motional state $\left|
1,0\right\rangle _{1}$ ($\left| 0,1\right\rangle _{1}$) the ion has a finite
probability of tunneling to the corresponding state of $\widetilde{V}_{2}$: $%
\left| 1,0\right\rangle _{2}$ ($\left| 0,1\right\rangle _{2}$). Therefore,
the motional basis states of the process are restricted to the ground and
first excited states of the harmonic traps, which is why we assumed from the
beginning that the parabolic approximation to the potential around each
minimum is designed to contain (at least) the two lowest states of the
harmonic oscillator. By collecting the resonance fluorescence signal in $%
\widetilde{V}_{1}$, we can probe the tunneling mechanism through the
behavior of the function ${\sf P}_{\downarrow }(t)$: the probability of
fluorescence measurement of the electronic ground state.

Considering the Lamb-Dicke limit and the fact that we have only one particle
in the process, the function $\exp \left( -\eta ^{2}\sum\nolimits_{n}%
\widehat{c}_{1}^{\left( n\right) ^{\dagger }}\widehat{c}_{1}^{\left(
n\right) }\right) $ can be fairly approximated by unity, so the Hamiltonian
governing the process has the simplified form (for $n=1,2)$%
\begin{eqnarray}
\widehat{{\cal H}} &=&{\cal R}^{\left( 1\right) }\left( \widehat{c}%
_{1}^{\left( 1\right) ^{\dagger }}\widehat{c}_{2}^{\left( 1\right) }+%
\widehat{c}_{1}^{\left( 1\right) }\widehat{c}_{2}^{\left( 1\right) ^{\dagger
}}\right) +{\cal R}^{\left( 2\right) }\left( \widehat{c}_{1}^{\left(
2\right) ^{\dagger }}\widehat{c}_{2}^{\left( 2\right) }+\widehat{c}%
_{1}^{\left( 2\right) }\widehat{c}_{2}^{\left( 2\right) ^{\dagger }}\right) 
\nonumber \\
&&+i\hbar \eta g\left( \sigma _{+}\widehat{c}_{1}^{\left( 1\right) ^{\dagger
}}\widehat{c}_{1}^{\left( 2\right) }e^{+i\varphi _{L}}-\sigma _{-}\widehat{c}%
_{1}^{\left( 1\right) }\widehat{c}_{1}^{\left( 2\right) ^{\dagger
}}e^{-i\varphi _{L}}\right) .  \label{Eq10}
\end{eqnarray}

Before proceeding further, it is worth mentioning that instead of adjusting
the ion-laser detuning to the first red sideband for the ion-laser
interaction ($\delta =\omega _{0}$), we could have chosen the carrier
interaction where $\delta =0$. In this case the effective Hamiltonian
following from Eq. (\ref{Eq8}) (with the above-mentioned approximations)
simplifies to 
\begin{eqnarray}
\widehat{{\cal H}}_{carrier} &=&{\cal R}^{\left( 1\right) }\left( \widehat{c}%
_{1}^{\left( 1\right) ^{\dagger }}\widehat{c}_{2}^{\left( 1\right) }+%
\widehat{c}_{1}^{\left( 1\right) }\widehat{c}_{2}^{\left( 1\right) ^{\dagger
}}\right) +{\cal R}^{\left( 2\right) }\left( \widehat{c}_{1}^{\left(
2\right) ^{\dagger }}\widehat{c}_{2}^{\left( 2\right) }+\widehat{c}%
_{1}^{\left( 2\right) }\widehat{c}_{2}^{\left( 2\right) ^{\dagger }}\right) 
\nonumber \\
&&+\hbar g\left( \sigma _{+}e^{+i\varphi _{L}}+\sigma _{-}e^{-i\varphi
_{L}}\right) {\rm {.}}  \label{Carrier}
\end{eqnarray}
We note that the Pauli operators act only when the ion is in harmonic well $%
1 $. The dynamics governed by Hamiltonians (\ref{Eq10}) and (\ref{Carrier})
make it possible to probe the tunneling mechanism of the ion by collecting
resonance fluorescence signals, as discussed above.

Starting from the initial state $\left| 1,0\right\rangle _{1}$ $\left|
0,0\right\rangle _{2}$ $\left| \uparrow \right\rangle $ and Hamiltonian (\ref%
{Eq10}) we obtain from the Schr\"{o}dinger evolution the evolved state 
\begin{eqnarray}
\left| \psi \left( t\right) \right\rangle &=&\left[ {\cal C}_{1}^{\left(
1\right) }\left( t\right) \left| 1,0\right\rangle _{1}\left|
0,0\right\rangle _{2}+{\cal C}_{2}^{\left( 1\right) }\left( t\right) \left|
0,0\right\rangle _{1}\left| 1,0\right\rangle _{2}\right] \left| \uparrow
\right\rangle  \nonumber \\
&&+\left[ {\cal C}_{1}^{\left( 2\right) }\left( t\right) \left|
0,1\right\rangle _{1}\left| 0,0\right\rangle _{2}+{\cal C}_{2}^{\left(
2\right) }\left( t\right) \left| 0,0\right\rangle _{1}\left|
0,1\right\rangle _{2}\right] \left| \downarrow \right\rangle ,  \label{Eq11}
\end{eqnarray}
whose coefficients satisfy the set of coupled linear equations 
\begin{mathletters}
\begin{eqnarray}
i\hbar \frac{d}{dt}{\cal C}_{1}^{\left( n\right) }\left( t\right) &=&{\cal R}%
^{\left( n\right) }{\cal C}_{2}^{\left( n\right) }\left( t\right)
-(-1)^{n}i\hbar \eta ge^{-(-1)^{n}i\varphi _{L}}{\cal C}_{1}^{\left(
m\right) }\left( t\right) ,  \label{Eq12a} \\
i\hbar \frac{d}{dt}{\cal C}_{2}^{\left( n\right) }\left( t\right) &=&{\cal R}%
^{\left( n\right) }{\cal C}_{1}^{\left( n\right) }\left( t\right) ,
\label{Eq12b}
\end{eqnarray}
with $n,m=1,2$ ($n\neq m$) and the initial conditions ${\cal C}_{1}^{\left(
1\right) }\left( 0\right) =1$, ${\cal C}_{2}^{\left( 1\right) }\left(
0\right) ={\cal C}_{1}^{\left( 2\right) }\left( 0\right) ={\cal C}%
_{2}^{\left( 2\right) }\left( 0\right) =0$. Now, we introduce a further
simplification into the present scheme. The larger the ratio $%
x_{0}^{2}/\Delta _{x}^{2}$ obtained by engineering the trap, the smaller
becomes the ratio of the matrix elements $\left| {\cal R}^{\left( 1\right) }/%
{\cal R}^{\left( 2\right) }\right| $. Therefore, when $\left| {\cal R}%
^{\left( 1\right) }/{\cal R}^{\left( 2\right) }\right| \ll 1$ we can neglect
the tunneling process between the ground states in the two harmonic wells.
In this case, the first term in Hamiltonian (\ref{Eq10}) can be disregarded
and, by adjusting the phase of the laser pulse such that $\varphi _{L}=-\pi
/2$, we obtain the coefficients: 
\end{mathletters}
\begin{mathletters}
\begin{eqnarray}
{\cal C}_{1}^{\left( 1\right) }\left( t\right) &=&\left[ \cos (\xi {\sf w}%
t)+\xi ^{2}-1\right] /\xi ^{2}{\rm {,\quad }}{\cal C}_{2}^{\left( 1\right)
}\left( 0\right) =0{\rm {,}}  \label{Eq13a} \\
{\cal C}_{1}^{\left( 2\right) }\left( t\right) &=&i\sin (\xi {\sf w}t)/\xi 
{\rm {,\quad }}{\cal C}_{2}^{\left( 2\right) }\left( 0\right) =\sqrt{\xi
^{2}-1}\left[ \cos (\xi {\sf w}t)-1\right] /\xi ^{2}{\rm {,}}  \label{Eq13b}
\end{eqnarray}
where the effective Rabi frequency, ${\sf w}=\eta g$, is modified by the
parameter $\xi =\left\{ 1+\left[ {\cal R}^{\left( 2\right) }/(\hbar \eta g)%
\right] ^{2}\right\} ^{1/2}$. Next we analyze the influence of parameter $%
\xi $ on the time-evolution of the probability of measuring fluorescence of
the electronic ground state ${\sf P}_{\downarrow }(t)=\left| {\cal C}%
_{1}^{\left( 2\right) }\left( t\right) \right| ^{2}$in harmonic well $1$.
Evidently, when ${\cal R}^{\left( 2\right) }=0$, we recover the well-known
dynamics of $P_{\downarrow }(t)$\ for JCH \cite{Wineland}. For the choice $%
{\cal R}^{\left( 2\right) }/(\hbar \eta g)=1$ ($\xi =\sqrt{2}$) and
employing the typical values in experiments with $^{40}$Ca$^{+}$ \cite%
{Blatt,Nagerl}{\Huge \ }$\eta \approx 0.1$, $g\approx 200$ kHz, and $\omega
_{0}\approx 2$ MHz (giving the estimate ${\cal R}^{\left( 2\right) }/(\hbar
\eta g)\approx 150\times {\cal R}^{\left( 2\right) }/E_{0}^{\left( 2\right)
})$, we obtain ${\cal R}^{\left( 2\right) }/E_{0}^{\left( 2\right) }\approx
7\times 10^{-3}$ as required to justify the approximation that the
superpositions $\phi _{\pm }^{n}(x)$ constitute eigenstates of the global
double-well potential. In fact, {\Huge \ }remembering that ${\cal R}^{\left(
2\right) }/(\hbar \eta g)=\left( \left. 3\omega _{0}x_{0}^{4}\right/ 16\eta
g\Delta _{x}^{4}\right) \exp \left( -x_{0}^{2}/2\Delta _{x}^{2}\right) $, we
obtain from the previous parameters the values $x_{0}^{2}/\Delta
_{x}^{2}\approx 17.3$ and $\left| {\cal R}^{\left( 1\right) }/{\cal R}%
^{\left( 2\right) }\right| \approx 6\times 10^{-2}$ which are in agreement
with the approximations considered above. In Fig. 1 we display the behavior
of function ${\sf P}_{\downarrow }(t)$ for ${\cal R}^{\left( 2\right) }=0$ ($%
\xi =1$, corresponding to the dynamics of the JCH) and ${\cal R}^{\left(
2\right) }/(\hbar \eta g)=1$. As anticipated by observing the expression for 
${\sf P}_{\downarrow }(t)$, the increase in the tunneling rate ${\cal R}%
^{\left( 2\right) }$ leads to an increase in the effective frequency $\xi 
{\sf w}$ of population inversion and, conversely, a decrease in the
amplitude of the oscillations of ${\sf P}_{\downarrow }(t)$, clearly
indicating the tunneling process. In fact, as soon as the ion reaches the
excited state of local well $1$, the coupling ${\cal R}^{\left( 2\right) }$
to local well $2$ entangles the motional excited states of both wells,
preventing the probability ${\sf P}_{\downarrow }(t)$ (associated with the\
measurement of state $\left| 0,1\right\rangle _{1}\left| \downarrow
\right\rangle $) from reaching unity. When the curve in Fig. 1 for ${\cal R}%
^{\left( 2\right) }/(\hbar \eta g)=1$ reaches its maxima (${\sf w}t=n\left.
\pi \sqrt{2}\right/ 4$, $n=1,2,...$), we have the entangled state $\frac{1}{2%
}\left| 1,0\right\rangle _{1}\left| 0,0\right\rangle _{2}\left| \uparrow
\right\rangle +\left( \frac{i}{\sqrt{2}}\left| 0,1\right\rangle _{1}\left|
0,0\right\rangle _{2}-\frac{1}{2}\left| 0,0\right\rangle _{1}\left|
0,1\right\rangle _{2}\right) \left| \downarrow \right\rangle $. Therefore,
both characteristics, the effective frequency $\xi {\sf w}$ of population
inversion and the amplitude of the oscillations of ${\sf P}_{\downarrow }(t)$%
, can be used to probe the single-particle tunneling mechanism.

It is worth observing that the choice of the alternative initial state $%
\left| 0,1\right\rangle _{1}$ $\left| 0,0\right\rangle _{2}$ $\left|
\downarrow \right\rangle $ leads to the entanglement $\left( \left|
1,0\right\rangle _{1}\left| 0,0\right\rangle _{2}\left| \uparrow
\right\rangle +\left| 0,0\right\rangle _{1}\left| 0,1\right\rangle
_{2}\left| \downarrow \right\rangle \right) /\sqrt{2}$. In this situation,
as expected for a closed system, we can re-establish a value of unity for $%
{\sf P}_{\downarrow }(t)$.

Now, turning to the Hamiltonian (\ref{Carrier}) and solving the Schr\"{o}%
dinger equation for the initial state $\left| 0,1\right\rangle _{1}$ $\left|
0,0\right\rangle _{2}$ $\left| \uparrow \right\rangle $, we obtain the
result 
\end{mathletters}
\begin{equation}
{\sf P}_{\downarrow }(t)=\frac{\left( \lambda _{1}\sin \left( \lambda
_{1}gt\right) -\lambda _{2}\sin \left( \lambda _{2}gt\right) \right) }{%
\left( \lambda _{2}^{2}-\lambda _{1}^{2}\right) ^{2}}^{2}{\rm {,}}
\label{Eq14}
\end{equation}
where the parameters modifying the Rabi frequency, $g$, are $\lambda _{i}=%
\frac{1}{\sqrt{2}}\left\{ 1+2\left[ {\cal R}^{\left( 2\right) }/(\hbar g)%
\right] ^{2}+\left( -1\right) ^{i}\left[ 1+4\left[ {\cal R}^{\left( 2\right)
}/(\hbar g)\right] ^{2}\right] ^{1/2}\right\} ^{1/2}$. Evidently, when $%
{\cal R}^{\left( 2\right) }=0$, we get the usual dynamics for the carrier
pulse, with ${\sf P}_{\downarrow }(t)=\left| \sin \left( gt\right) \right|
^{2}$. In Fig. 2 we display the time evolution of Eq. (\ref{Eq14}), for $%
{\cal R}^{\left( 2\right) }=0$ and ${\cal R}^{\left( 2\right) }/(\hbar g)=1$ 
$\left( \lambda _{i}=\left[ \left( 3+(-1)^{i}\sqrt{5}\right) /2\right]
^{1/2}\right) $. For the typical values given above, we obtain ${\cal R}%
^{\left( 2\right) }/E_{0}^{\left( 2\right) }\approx 7\times 10^{-2}$ (a
value which match $x_{0}^{2}/\Delta _{x}^{2}\approx 10.8$ and $\left| {\cal R%
}^{\left( 1\right) }/{\cal R}^{\left( 2\right) }\right| \approx 0.1$).
Similarly to the behavior displayed in Fig. 1, we observe that the frequency
of population inversion is higher than the Rabi frequency $g$ associated
with free carrier dynamics. However, differently from the situation in
Fig.1, the probability ${\sf P}_{\downarrow }(t)$ (displaying the
characteristic beat pattern due to Eq. (\ref{Eq14})) can still reach unity.
Since in the carrier regime the electronic states do not couple to the
motional states, the excited and ground electronic states are both subject
to the tunneling process (differently from the JCH case), resulting in the
interference pattern shown in Fig. 2.

From the value for the position uncertainty in a harmonic oscillator ground
state $\Delta _{x}^{2}=\hbar /\left( 2m\omega _{0}\right) $ computed for $%
^{40}$Ca$^{+}$ with motional frequency $\omega _{0}\approx 2$ MHz, and
considering the Jaynes-Cummings regime, where $x_{0}^{2}/\Delta
_{x}^{2}\approx 17.3$, we obtain (using the typical values $\eta \approx 0.1$
and $g\approx 200$ kHz \cite{Blatt,Nagerl}) the distance between the local
minima $2x_{0}\approx 0.16$ $\mu $m.\ Making the same estimates for the
Carrier regime, where $x_{0}^{2}/\Delta _{x}^{2}\approx 10.3$, we obtain the
distance between the local minima $2x_{0}\approx 0.13$ $\mu $m. Recently, it
was reported in Ref. \cite{Nagerl} the laser addressing of individual ions
in a linear ion trap with frequency $125$ kHz when the distance between the
ions is about $19.\mu $m. The authors of Ref. \cite{Nagerl} argue that the
addressing technique presented permits individual addressing when the
distance between the ions is only $7.6.\mu $m with small error. Therefore,
with nowadays technology it is difficult to individually address each local
potential well. On the other hand, when considering the situation $b)$ where
the laser beams reach the local wells simultaneously, the Hamiltonian \ref%
{Eq10} (in Jaynes-Cummings regime and assuming $\left| {\cal R}^{\left(
1\right) }/{\cal R}^{\left( 2\right) }\right| \ll 1$) turns to be

\begin{eqnarray}
\widehat{{\cal H}} &=&{\cal R}^{\left( 2\right) }\left( \widehat{c}%
_{1}^{\left( 2\right) ^{\dagger }}\widehat{c}_{2}^{\left( 2\right) }+%
\widehat{c}_{1}^{\left( 2\right) }\widehat{c}_{2}^{\left( 2\right) ^{\dagger
}}\right)  \nonumber \\
&&+i\hbar \eta g\left[ \sigma _{+}\left( \widehat{c}_{1}^{\left( 1\right)
^{\dagger }}\widehat{c}_{1}^{\left( 2\right) }+\widehat{c}_{2}^{\left(
1\right) ^{\dagger }}\widehat{c}_{2}^{\left( 2\right) }\right) e^{+i\varphi
_{L}}\right. .  \nonumber \\
&&-\left. \sigma _{-}\left( \widehat{c}_{1}^{\left( 1\right) }\widehat{c}%
_{1}^{\left( 2\right) ^{\dagger }}+\widehat{c}_{2}^{\left( 1\right) }%
\widehat{c}_{2}^{\left( 2\right) ^{\dagger }}\right) e^{-i\varphi _{L}}%
\right] .  \label{Eq15}
\end{eqnarray}
Starting from the initial state $\left| 1,0\right\rangle _{1}$ $\left|
0,0\right\rangle _{2}$ $\left| \uparrow \right\rangle $ and Hamiltonian (\ref%
{Eq15}) we obtain from the Schr\"{o}dinger evolution the evolved state
described by Eq. (\ref{Eq11}) with the coefficients satisfying the coupled
linear equations

\begin{mathletters}
\begin{eqnarray}
i\hbar \frac{d}{dt}{\cal C}_{n}^{\left( 1\right) } &=&i\hbar ge^{i\varphi
_{L}}{\cal C}_{n}^{\left( 2\right) }\left( t\right) ,  \label{Eq16a} \\
i\hbar \frac{d}{dt}{\cal C}_{n}^{\left( 2\right) }\left( t\right) &=&{\cal R}%
^{\left( 2\right) }{\cal C}_{m}^{\left( 2\right) }\left( t\right) -i\hbar
ge^{-i\varphi _{L}}{\cal C}_{n}^{\left( 1\right) }\left( t\right) ,
\label{Eq16b}
\end{eqnarray}
with $n,m=1,2$ ($n\neq m$) and the initial conditions ${\cal C}_{1}^{\left(
1\right) }\left( 0\right) =1$, ${\cal C}_{2}^{\left( 1\right) }\left(
0\right) ={\cal C}_{1}^{\left( 2\right) }\left( 0\right) ={\cal C}%
_{2}^{\left( 2\right) }\left( 0\right) =0$. Solving the system above we
obtain the coefficients:

\end{mathletters}
\begin{mathletters}
\label{0}
\begin{eqnarray}
{\cal C}_{1}^{\left( 1\right) }\left( t\right) &=&\frac{1}{(\chi ^{2}+4)}%
\sum_{j=1,2}\left[ \chi ^{2}-\left( \lambda _{j}\right) ^{2}+3\right] \cos
\left( \lambda _{j}gt\right) {\rm {,}}  \label{Eq17a} \\
{\cal C}_{2}^{\left( 1\right) }\left( t\right) &=&\frac{i}{\chi (\chi ^{2}+4)%
}\sum_{j=1,2}\lambda _{j}\left[ \left( \chi ^{2}+2\right) ^{2}-\left( \chi
^{2}+2\right) \left( \lambda _{j}\right) ^{2}\right] \sin \left( \lambda
_{j}gt\right) {\rm {,}}  \label{Eq17b} \\
{\cal C}_{1}^{\left( 2\right) }\left( t\right) &=&\frac{e^{-i\varphi _{L}}}{%
(\chi ^{2}+4)}\sum_{j=1,2}\lambda _{j}\left[ \left( \lambda _{j}\right)
^{2}-\chi ^{2}-3\right] \sin \left( \lambda _{j}gt\right) {\rm {,}}
\label{Eq17c} \\
{\cal C}_{2}^{\left( 2\right) }\left( t\right) &=&\frac{-ie^{-i\varphi _{L}}%
}{\chi (\chi ^{2}+4)}\sum_{j=1,2}\left[ 2\left( \lambda _{j}\right)
^{2}-\chi ^{2}-2\right] \cos \left( \lambda _{j}gt\right) {\rm {,}}
\label{Eq17d}
\end{eqnarray}
where $\lambda _{j}=\left\{ 1+\left. \left[ \chi ^{2}+\left( -1\right)
^{j}\chi \sqrt{4+\chi ^{2}}\right] \right/ 2\right\} ^{1/2}$ and $\chi
=\left. {\cal R}^{\left( 2\right) }\right/ \hbar \eta g$. The probability of
measuring the electronic ground state $\left| \downarrow \right\rangle $ of
the ion in both local wells $1$ and $2$ is given by ${\sf P}_{\downarrow
}(t)=\left| {\cal C}_{1}^{\left( 2\right) }\left( t\right) +{\cal C}%
_{2}^{\left( 2\right) }\left( t\right) \right| ^{2}$. In Fig. 3 we display
the behavior of function ${\sf P}_{\downarrow }(t)$ for ${\cal R}^{\left(
2\right) }=0$ (corresponding to the well know Jaynes-Cummings dynamics) and $%
{\cal R}^{\left( 2\right) }/(\hbar \eta g)=1$ ( $\lambda _{1}=\sqrt{\left. 3-%
\sqrt{5}\right/ 2}$ and $\lambda _{2}=\sqrt{\left. 3+\sqrt{5}\right/ 2}$).
Similarly to the behavior displayed in Fig. 1, we observe that the frequency
of population inversion in Fig. 3 (full line) is higher than the Rabi
frequency $g$ associated with the free JCH dynamics whereas the amplitude of
the oscillations of ${\sf P}_{\downarrow }(t)$ is smaller than that for the
free JCH dynamics. Both characteristics clearly indicate the tunneling
process in Fig. 3, without the requirement of the individual addressing of
each local potential well.

In conclusion, we have presented a scheme for probing the tunneling process
of a single ion trapped in a double-well electromagnetic potential. The
tunneling dynamics is characterized by the behavior of the probability of
collecting an electronic ground state fluorescence signal in one of the
local potential wells. Two situations were analyzed, when collecting
fluorescence measurement $a)$ by individual addressing the ion in harmonic
well $1$, and $b)$ considering the laser beams to reach the local wells
simultaneously. Although the situation $a)$ is beyond nowadays technology,
the single-particle quantum tunneling can clearly be probed through the
situation $b)$. Considering the possibility of individual addressing of
local wells, the present proposal can be extended to the implementation of
the fundamental controlled-NOT two-bit gate by adding another laser beam in
harmonic well $2$\ and storing the quantum bits in the motional states of
both local wells \cite{Roberto}. We also note that our proposal can be
implemented in spite of the decoherence processes coming from the coupling
of the motional modes with the residual background gas \cite{Serra} and with
classical stochastic electric field \cite{Matos}, in addition to\ the finite
lifetime of the electronic levels\cite{Matos,Vogel}. From the experimental
results reported in \cite{Wineland,Blatt} we observe that several Rabi
oscillations are visible within the decoherence time making it possible to
observe the signature of the tunneling process. However, we stress that the
designing of the double-well potential would require that the trap
electrodes be only few micron away from the ion; then it is likely that the
heating rate in the trap becomes higher than that observe in single traps,
increasing the decoherence rate \cite{Roos}. Besides, the required trap
structure makes difficult to obtain high trap frequencies, about MHz, and
consequently, makes difficult the cooling process of the ion to its motional
ground state \cite{Roos}. Here we note that the cooling process could be
implemented in a single trap structure which could be adiabatically modified
to a double-well potential.

Although the design of such a double-well trap may turn out to be a
considerable technical challenge, the fundamental principles discussed here
can be implemented by engineering two-mode interactions in ion traps \cite%
{Knight}, in which the motional states of the ion in two different
directions can be coupled, analogously to the dynamics for the double-well.
Besides, the proposal here presented might provide a motivation for future
experimental work.

\bigskip

{\bf Acknowledgments}

We wish to express thanks for the support from FAPESP (under contracts
\#99/11617-0, \#00/15084-5, and \#02/02633-6) and CNPq (Intituto do Mil\^{e}%
nio de Informa\c{c}\~{a}o Qu\^{a}ntica), Brazilian agencies. We also thank
Christian Roos for helpful discussions.

\bigskip

{\bf Figure Captions}

\bigskip

Figure 1. The evolution of function ${\sf P}_{\downarrow }(t)$, in the
Jaynes-Cumming regime and individual addressing of local well $1$, for the
values ${\cal R}^{\left( 2\right) }/(\hbar \eta g)=0$ (dotted line) and $1$
(full line).

Figure 2. The evolution of function ${\sf P}_{\downarrow }(t)$, in the
carrier regime and individual addressing of local well $1$, for the values $%
{\cal R}^{\left( 2\right) }/(\hbar g)=0$ (dotted line) and $1$ (full line).

Figure 3. The evolution of function ${\sf P}_{\downarrow }(t)$, in the
Jaynes-Cumming regime and the laser beams reaching both local wells
simultaneously, for the values ${\cal R}^{\left( 2\right) }/(\hbar \eta g)=0$
(dotted line) and $1$ (full line).
\end{mathletters}


\bigskip

\begin{references}
\bibitem{comm} J. I. Cirac, P. Zoller, H. J. Kimble, and H. Mabuchi, Phys.
Rev. Lett. {\bf 78}, 3221 (1997); T. Pellizzari, {\it ibid}. {\bf 79}, 5242
(1997).

\bibitem{comp} J. I. Cirac and P. Zoller, Phys. Rev. Lett. {\bf 74}, 4091
(1995); Q. A. Turchette, C. J. Hood, W. Lange, H. Mabuchi, and H. J. Kimble, 
{\it ibid.} {\bf 75}, 4710 (1995); I. L. Chuang, L. M. K. Vandersypen , X.
L. Zhou, D. W. Leung, S. Lloyd, Nature, {\bf 393}, 143 (1998); B. E. Kane, 
{\it ibid}. {\bf 393}, 143 (1998); L. M. K. Vandersypen, M. Steffen, G.
Breyta, C. S. Yannoni, M. H. Sherwood, I. L. Chuang, {\it ibid}. {\bf 414},
883 (2001)

\bibitem{EPR} A. Einstein, B. Podolsky, and N. Rosen, Phys. Rev. {\bf 47},
777 (1935).

\bibitem{CL} A. O. Caldeira and A. J. Leggett, Phys. Rev. Lett. {\bf 46},
211 (1981).

\bibitem{Leggett} A. J. Leggett, S. Chakravarty, A. T. Dorsey, Matthew P. A.
Fisher, Anupam Garg, and W. Zwerger, Rev. Mod. Phys. {\bf 59}, 1 (1987).

\bibitem{Milburn} G. J. Milburn, J. Corney, E. M. Wright, and D. F. Walls,
Phys. Rev. A {\bf 55}, 4318 (1997); J. F. Corney and G. J. Milburn, {\it ibid%
}. {\bf 58}, 2399 (1998).

\bibitem{Ketterle} M. R. Andrews, C. G. Townsend, H. J. Miesner, D. S.
Durfee, D. M. Kurn, and W. Ketterle, Science, {\bf 275}, 637 (1997).

\bibitem{Wineland} D. J. Wineland, C. Monroe, W. M. Itano, D. Leibfried, B.
E. King, and D. M. Meekhof, J. Res. NIST {\bf 103}, 259 (1998).

\bibitem{Blatt} Ch. Roos, Th. Zeiger, H. Rohde, H. C. N\"{a}gerl, J.
Eschner, D. Leibfried, F. Schmidt-Kaler, and R. Blatt, Phys. Rev. Lett. {\bf %
83}, 4713 (1999).

\bibitem{Nagerl} H. C. N\"{a}gerl, D. Leibfried, H. Rohde, G. Thalhammer, J.
Eschner, F. Schmidt-Kaler, and R. Blatt, Phys. Rev. A {\bf 60}, 145 (1999).

\bibitem{Roberto} R. M. Serra {\it et al.} To be published elsewhere.

\bibitem{Serra} R. M. Serra, N. G. de Almeida, W. B. da Costa, and M. H. Y.
Moussa,{\it \ }Phys. Rev. A {\bf 64}, 033419 (2001).

\bibitem{Matos} A. A. Budini, R. L. de Matos Filho, and N. Zagury, Phys.
Rev. A {\bf 65}, 041402 (2002).

\bibitem{Vogel} C. Di Fidio and W. Vogel, Phys. Rev. A {\bf 62}, 031802(R)
(2000).

\bibitem{Roos} Ch. Roos, private communication.

\bibitem{Knight} J. Steinbach, J. Twamley, and P. L. Knight {\it ,} Phys.
Rev. A {\bf 56}, 4815 (1997).
\end{references}
\end{document}